\documentclass[conference]{IEEEtran}
\IEEEoverridecommandlockouts
\let\svtilde~

\newcommand\newtildeON[1][A]{\def~{\csname newtilde#1\endcsname}}
\newcommand\newtildeOFF{\let~\svtilde}
\usepackage{cite}

\usepackage{amsmath,amssymb,amsfonts}
\usepackage{algorithmic}
\usepackage{graphicx}
\usepackage{textcomp}
\usepackage{listings}
\usepackage[table]{xcolor}
\usepackage{mathtools}
\usepackage{diagbox}
\usepackage{enumitem} 
\usepackage{subfigure}
\definecolor{codegreen}{rgb}{0,0.6,0}
\definecolor{codegray}{rgb}{0.5,0.5,0.5}
\definecolor{codepurple}{rgb}{0.58,0,0.82}
\definecolor{backcolour}{rgb}{0.95,0.95,0.92}

\definecolor{dkgreen}{rgb}{0,0.6,0}
\definecolor{gray}{rgb}{0.5,0.5,0.5}
\definecolor{mauve}{rgb}{0.58,0,0.82}
\def\BibTeX{{\rm B\kern-.05em{\sc i\kern-.025em b}\kern-.08em
    T\kern-.1667em\lower.7ex\hbox{E}\kern-.125emX}}
\begin{document}

\title{DareFightingICE Competition: A Fighting Game Sound Design and AI Competition
\\

}

\author{\IEEEauthorblockN{Ibrahim Khan, Thai Van Nguyen, Xincheng Dai}
\IEEEauthorblockA{\textit{Graduate School of Information Science and Engineering} \\
\textit{Ritsumeikan University}\\
Kusatsu, Shiga, Japan \\
\{gr0556vx, gr0557fv, gr0502pv\}@ed.ritsumei.ac.jp}
\and
\IEEEauthorblockN{Ruck Thawonmas}
\IEEEauthorblockA{\textit{College of Information Science and Engineering} \\
\textit{Ritsumeikan University}\\
Kusatsu, Shiga, Japan \\
ruck@is.ritsumei.ac.jp}

}

\maketitle

\begin{abstract}
This paper presents a new competition -- at the 2022 IEEE Conference on Games (CoG) -- called DareFightingICE Competition. The competition has two tracks: a sound design track and an AI track.
The game platform for this competition is also called DareFightingICE, a fighting game platform. DareFightingICE is a sound-design-enhanced version of FightingICE, used earlier in a competition at CoG until 2021 to promote artificial intelligence (AI) research in fighting games. In the sound design track, participants compete for the best sound design, given the default sound design of DareFightingICE as a sample, where we define a sound design as a set of sound effects combined with the source code that implements their timing-control algorithm. Participants of the AI track are asked to develop their AI algorithm that controls a character given only sound as the input (blind AI) to fight against their opponent; a sample deep-learning blind AI will be provided by us.
Our means to maximize the synergy between the two tracks are also described. This competition serves to come up with effective sound designs for visually impaired players, a group in the gaming community which has been mostly ignored. To the best of our knowledge, DareFightingICE Competition is the first of its kind within and outside of CoG.
\end{abstract}

\begin{IEEEkeywords}
Sound  Design Competition, AI  Competition, Visually  Impaired  Players,  DareFightingICE, FightingICE, Fighting  Game
\end{IEEEkeywords}

\section{Introduction}
Video games are on the rise, with the global market revenue of video games being increased from 178.37 million to 197.11 million (U.S dollar) from 2021 to 2022 \cite{b1}. Their popularity is increasing at a fast and steady pace. Since the start, video games have been a testing ground for Artificial Intelligence (AI) \cite{b2}. There have been many breakthroughs in AI stemming from the research on video games \cite{b3}. The rise in the popularity of games is the reason that game developers are looking to expand their audience. One of which is visually impaired players (VI Players), which have been mostly ignored in the past \cite{b4}. Game developers or researchers are adding new features such as specific audio cues so that VI Players can also experience and enjoy the games \cite{b5}.\par

Our focus is on fighting games. In these fighting games, players go against another player or a computer player in a one-versus-one fight, using different attacks and abilities to overcome the opponent. These fighting games are mostly two-dimensional, which means that the players can only move in two dimensions. In addition, because fighting games are fairly simple compared to other genres of video games when they are given the right sound designs, VI Players will have a much easier time playing than other genres; in this work, we define a sound design as a set of sound effects combined with the source code that implements their timing-control algorithm.

Our goal is to use a fighting game platform called “DareFightingICE”, which is “FightingICE” \cite{b6} with an enhanced sound design, to help VI Players play the game and to set the DareFightingICE platform as a testing ground for AI algorithms that only use sound as the input. We plan to achieve this goal through running a competition in the upcoming IEEE Conference on Games (CoG). This competition has two tracks: one is a sound design competition, and the other is an AI competition that only uses sound as the input.

We anticipate that this research will give voice to one of the most ignored members of the gaming community -- VI Players -- and that we hope the competition will give hints to the production of sound designs that can be used in video games to make them compatible with VI Players. Along with the sound design, this competition is going to open doors to a new wave of blind AI algorithms, i.e., those that only use sound as the input. Our contributions are as follows: first, we provide a platform for developers and researchers to create sound designs considering VI Players for fighting games; second, we are opening the door for research in AI algorithms that only use sound as the input; and third, the two tracks -- Sound Design and AI -- in this competition can be expected to have synergy between respective research areas.

\section{Related Work}
Since our research is a combination of three different topics (sound designs in video games, games for VI Players, and competitions), this section is divided into three parts.

\subsection{Sound/ Sound Designs in Video Games}

There has been plenty of research on the importance of sound in video games. It has been established that music played in video games can affect a person’s daily life, personality, and such \cite{b7}. These effects can be formed because of a person’s attachment to the music used or played in a game. Sound in video games in terms of voiceover (audio dialogue played by the player or the non-playable characters) has been known to be more engaging for the players \cite{b8}. These voiceovers also help in remembering the information provided to the players in the game. The background music being played in the game also has an important role to play. It was observed that players performed better with background music than those who played the game without background music \cite{b9}.\par

Similarly, it has been observed that the music played in video games must meet the mood or tone of the game to be engaging for the players \cite{b10}. A bad example of this would be to play calm music when the game is at a climax. Moving from general music to specific sound effects and audio cues, it has been seen that directional or 3D sound effects help the players know more about their environment, such as knowing the position of other players \cite{b11}. These 3D sound effects include sound like footsteps and other movement sound. \par

So far, this sub-section has been about sound in video games. Now shifting from sound in video games to sound designs, there has been only a little research when it comes to sound designs in video games. And even fewer have had positive results that can be useful in the creation of a sound design. Scarce as the research may be, sound designs have been proposed for game genres such as First Person Shooter and Real-Time Strategy \cite{b12}. These two sound designs have specific sound effects for respective game genres. 

\subsection{Games for VI Players and Fighting Games}
This sub-section includes research related to VI Players and fighting games that have some sort of support for such players. The reason behind the inclusion of fighting games here is because the main platform being used for our research is a fighting game platform and thereby our focus is on the fighting game genre of video games. Before making games for VI Players, it is important to know what these players want in a game and what they look forward to. Keeping this in mind it was observed that fighting games are the second most played games among VI Players \cite{b13}. There has been some research on serious games (a game with a purpose like an educational game) which involve VI Players. It was observed that audio cues in a serious game were able to convey spatial information about a room to such players \cite{b14}. These audio cues were played as VI Players explored a room in the game.

Moving from research on games for VI Players to fighting games with VI Players compatibility i.e., fighting games that enable the VI player to play with their sound effects. In this work, we focus on three fighting games with such compatibility, to the best of our knowledge. The first, but the least compatible among the three, is Takken 7, the latest game in the game series called Takken. This game has specific sound for different characters making it easy to recognize and remember. The next is Mortal Kombat 11, the latest game in the fighting game series Mortal Kombat. This game is even more helpful in terms of VI Players compatibility than Takken 7 because not only does it have specific sound for different characters, but it also gives different audio cues for different characters on the character selection screen. Lastly, Killer Instinct is a fighting game that has stereo sound and different sound effects, such as a sound effect played when the player reaches the end of the stage, and a sound effect indicating the increase in energy, to help VI Players play the game. Different concepts of audio cues in this game are adopted in DareFightingICE.\par

All the aforementioned commercial games can be played by VI Players, but there is room for improvement. The Tekken and Mortal Kombat series do not have 3D sound, and Killer Instinct’s 3D sound can get VI Players overwhelmed by the different sound effects of special moves a character can perform, which are loud. In addition, these games are somewhat not user-friendly for VI Players, as they were not specifically designed for VI Players or at least their publishers did not claim so. Our new fighting game platform is created to solve all these problems.\par

\subsection{Competitions and Sound Design Competitions}
This sub-section is about different competitions regarding video games at CoG as well as sound design competitions.
CoG has many competitions. Those with recent papers at CoG include Hanabi Competition \cite{b15} which is based on the game Hanabi, a cooperative card game; Blood Bowl Competition \cite{b16} which is a board game competition for AI; VGC AI Competition \cite{b17} which is an AI competition based on Pokémon; and Carle’s Game AI competition \cite{b18} which is a challenge in open-ended machine exploration and creativity.

Now let us shift from competitions in CoG to sound design competitions outside of CoG. This is because so far there has been no sound design competition in CoG. There have been a few sound design competitions conducted in the past. Game Music and Sound Design Conference (GMSDC’18) is a sound design competition for which the participants are allowed to submit a sound design for all genres of games \cite{b19}. The Berlin International Sound Design Competition (BISDC’18) is a sound design competition for game developers and filmmakers, allowing sound designs for all genres of games and films \cite{b20}. Film Music Contest (FMC)’s Music for Video Game (MVG’21) is a sound design competition for all types of games, just like GMSDC'18 \cite{b21}. These three sound design competitions are conducted outside of any academic conference related to games. \par

\begin{table}[t!]
    \caption{Comparison of existing competitions and our competition over four different contribution points.
(A)	Normal AI Contribution (B) Sound Design Contribution (C) Blind AI Contribution (D) Sound Design for VI Players
}
    
    \label{Comp-Comparison}
    \begin{center}
    \begin{tabular}{|c|c|c|c|c|}
    \hline
    \textbf{Competition}&\multicolumn{4}{|c|}{\textbf{Contributions}} \\
    \cline{2-5} 
    \textbf{Names} & \textbf{\textit{A}}& \textbf{\textit{B}}& \textbf{\textit{C}} & \textbf{\textit{D}} \\
    \hline
    Bot Bowl III& $\surd$ & $\times$ & $\times$ & $\times$ \\
    \hline
    ColorShapeLinks& $\surd$ & $\times$ & $\times$ & $\times$ \\
    \hline
    Dota 2 5v5 AI Competition& $\surd$ & $\times$ & $\times$ & $\times$ \\
    \hline
    Fighting Game AI Competition& $\surd$ & $\times$ & $\times$ & $\times$ \\
    \hline
    GVGAI single-player learning competition& $\surd$ & $\times$ & $\times$ & $\times$ \\
    \hline
    Ludii AI Competition& $\surd$ & $\times$ & $\times$ & $\times$ \\
    \hline
    StarCraft AI Competition& $\surd$ & $\times$ & $\times$ & $\times$ \\
    \hline
    Strategy Card Game AI Competition& $\surd$ & $\times$ & $\times$ & $\times$ \\
    \hline
    AI Snakes Game& $\surd$ & $\times$ & $\times$ & $\times$ \\
    \hline
    GMSDC’18& $\times$ & $\surd$ & $\times$ & $\times$ \\
    \hline
    BISDC’18& $\times$ & $\surd$ & $\times$ & $\times$ \\
    \hline
    MVG’21& $\times$ & $\surd$ & $\times$ & $\times$ \\
    \hline
    Our Competition& $\times$ & $\times$ & $\surd$ & $\surd$ \\
    \hline
    \multicolumn{5}{l}{$\surd$ : Yes , $\times$ : No} \\

    \end{tabular}
    \end{center}
    
\end{table}




Table~\ref{Comp-Comparison} summarizes the contributions of nine competitions in CoG 21 and the aforementioned three sound design competitions. To the best of our knowledge, those nine were the competitions at CoG 21 that had participants, while the other three were the only prominent sound design competitions in recent years. The definition of each contribution in the table is as follows: 

\begin{description}
\item[Normal AI:] Targeting AI algorithms that can access game state data in all kinds of forms.
\item[Sound Design:] Targeting general sound designs for any field or game, with no requirements or restrictions on what should be included.
\item[Blind AI:] Targeting AI algorithms that only use sound as the input.
\item[Sound Design for VI Players:] Targeting sound designs for VI Players.
\end{description}

This table should manifest well the position of our competition among the rest. Note that the nine CoG 21 competitions did not provide audio data to their entry AIs.

\section{DareFightingICE}
\label{DareFightingICE}

Compared to its predecessor FightingICE, DareFightingICE has an enhanced sound design, which is made by keeping VI Players in mind. It uses 3D sound effects. Along with the enhanced sound design, DareFightingICE also has a new interface, enabling AIs (henceforth AI in a countable form refers to an AI algorithm) to receive sound data from the game.\par

The FightingICE platform supports Java programming and is also wrapped for Python programming by Py4J to develop AIs. This platform allows users to create AIs for non-player characters. The platform is also useful for relevant research communities to conduct research as shown in the following paragraph.\par

In one paper, their authors presented a fighting-game gameplay generation using Monte-Carlo tree search with the evaluation function formed based on highlight cues \cite{b22}. In another paper, Chen et al. \cite{b23} proposed a method, using FightingICE, to learn utility functions from data collected from human players, to recreate a target behavior. Another paper \cite{b24} proposed an algorithm that combines Rolling Horizon Evolution Algorithm with an opponent model using FightingICE; their AI won the 2020 Fighting Game AI Competition. These three are just examples of recent studies utilizing the FightingICE platform. \par

As with the FightingICE platform, we hope that the DareFightingICE platform is going to open exciting new doors for developers and researchers. With the addition of blind AIs and sound designs for VI Players, DareFightingICE is the first of its kind, as a research platform, for fighting games. We hope to see DareFightingICE being used in research by relevant research and development communities in the near future.

\subsection{Sample Sound Design}

Our sample sound design is created and used as the default sound design of DareFightingICE to provide the participants of this competition with a sample of how a sound design can be created. In its current form, there is room for improvement, which is intentional, and it is our hope to see improved and stimulating new sound designs. The sample sound design is created with the java library "lwjgl", which uses openAL. \par

There is a total of 51 sound effects in the sample sound design including the background music. Some sound effects are the same for similar moves or actions. Since we use openAL, the sound effects are mono, not stereo. This is because to emit 3D sound with openAL the sound effects must be in mono. Of the 51 sound effects, there are three special sound effects (SSEs): Heartbeat, Energy-Increase, and Border-Alert. These SSEs are to help VI Players play the game. Energy-Increase and Border-Alert SSEs are inspired by Killer Instinct, and Heartbeat is from previous research on sound design \cite{b12}; their descriptions are as follows:
Heartbeat: This sound effect is played when the player’s health is below 50. For player 1 the sound effect is played on the left speaker and for player 2 on the right.
Energy Increase: This sound effect is played when the player’s energy goes +50 from the previous value. For player 1 the sound effect is played on the left speaker and for player 2 on the right.
Border Alert: This sound is played when a player reaches the end of the stage on either side. The sound is played on the left side when a player reaches the end of the stage on the left side and the same applies for the right side. Please also use the description environment.\par

There are three important elements in our design: Listener, Audio-Sources, and Audio-Buffers. Listener, provided by lwjgl, listens to the audio generated by the game. It is responsible for providing the output of audio to the game players. The position of Listener in DareFightingICE is in the middle of the game screen since the camera does not move from its original position. The velocity of Listener is defined by its speed and direction, which is used to create a Doppler effect. The orientation of Listener is represented by two vectors: ``at", and ``up". The former represents the forward direction, and the latter represents the upward direction. \par

Audio-Sources are in-game objects or in-game events that produce sound, and in our sample sound design, the position of these Audio-Sources from Listener determines from which direction the players will hear the sound effect. These Audio-Sources also have velocity and orientation. A Doppler effect is created with the help of the velocity of an Audio-Source of interest and the velocity of Listener. Audio-Buffers are all the sound effects that our sample sound design has. Figure \ref{fCode3} shows the provided function to play a sound effect. In this figure, \textit{soundRenderers} contains two renderers (one for rendering sound for the speakers and the other one for AIs), \textit{bufferId} represents the index of an Audio-Buffer, the parameters \textit{x} and \textit{y} store the horizontal location and the vertical locations of where to play the sound effect on the stage (map), and \textit{loop} confirms whether to play the sound effect in a loop or not. \par

\begin{figure}[t!]
\begin{lstlisting}
    /**
    * Plays an Audio-Source by giving it an Audio-Buffer to play and the position where to play.
    */
    public void play2(AudioSource source, AudioBuffer buffer, int x, int y, boolean loop){
        for (int i = 0; i < soundRenderers.size(); i++){
            int sourceId = source.getSourceIds()[i];
            int bufferId = buffer.getBuffers()[i];
            soundRenderers.get(i).play(sourceId, bufferId, x, y, loop);
        }
    }
\end{lstlisting}
\caption{Function to play a sound effect}
\label{fCode3}
\end{figure}

\subsection{AI Interface and Blind AIs}
Below are details about our AI interface that provides sound data to blind AIs and is used in the competition. \par

There have been a number of prior studies relating to game playing AIs with sound. Gaina and Stephenson\cite{b5} expanded the General Video Game AI framework to support sound and trained an AI that played the game from sound only. Hegde et al.\cite{b25} extended the VizDoom framework to provide the in-game sound to AIs and trained them in several scenarios with increasing difficulty to test the perception of sound. Results from these studies show promise in learning to play games from sound. In our project, sound in DareFightingICE will be provided to entry blind AIs so they can be trained to play the game from sound only.

To facilitate AIs with access to sound data, the original interface in FightingICE\cite{b6} is extended with a new method \textit{getAudioData}() that provides sound data at each frame of the game. At each frame, sound data are sampled with a length of 16.67 ms; note that DareFightingICE has an FPS value of 60, as in FightingICE. In DareFightingICE, stereo sound is provided in 3D format, which helps effectively represent the game's environment. Therefore, in order to allow AIs' perception of sound that might help them localize in-game objects and in-game events, a two-channel format of sound is provided. 

\begin{figure}[t!]
\centerline{\includegraphics[width=0.4\textwidth]{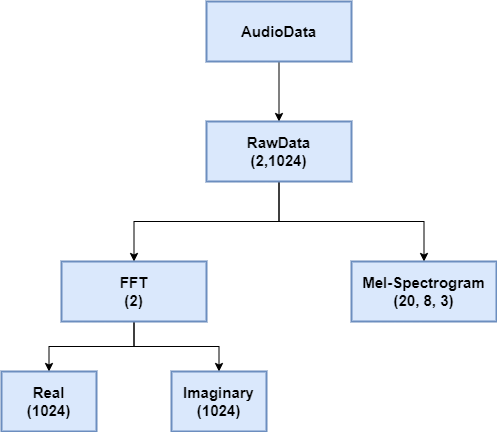}}
\caption{Sound data structure.}
\label{figSoundDataStructure}
\end{figure}

\begin{figure*}
    \centering
    \includegraphics[width=.33\linewidth]{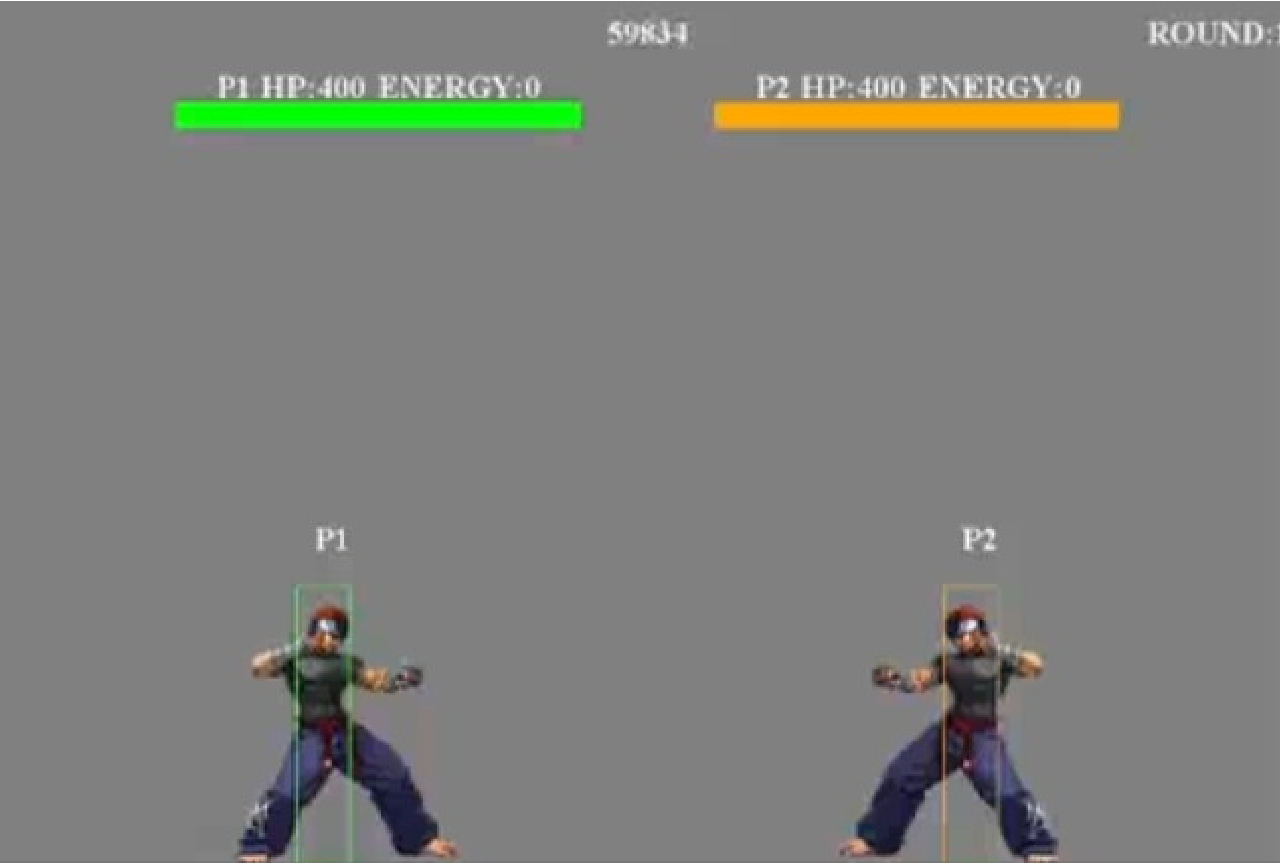}\hfill
    \includegraphics[width=.33\linewidth]{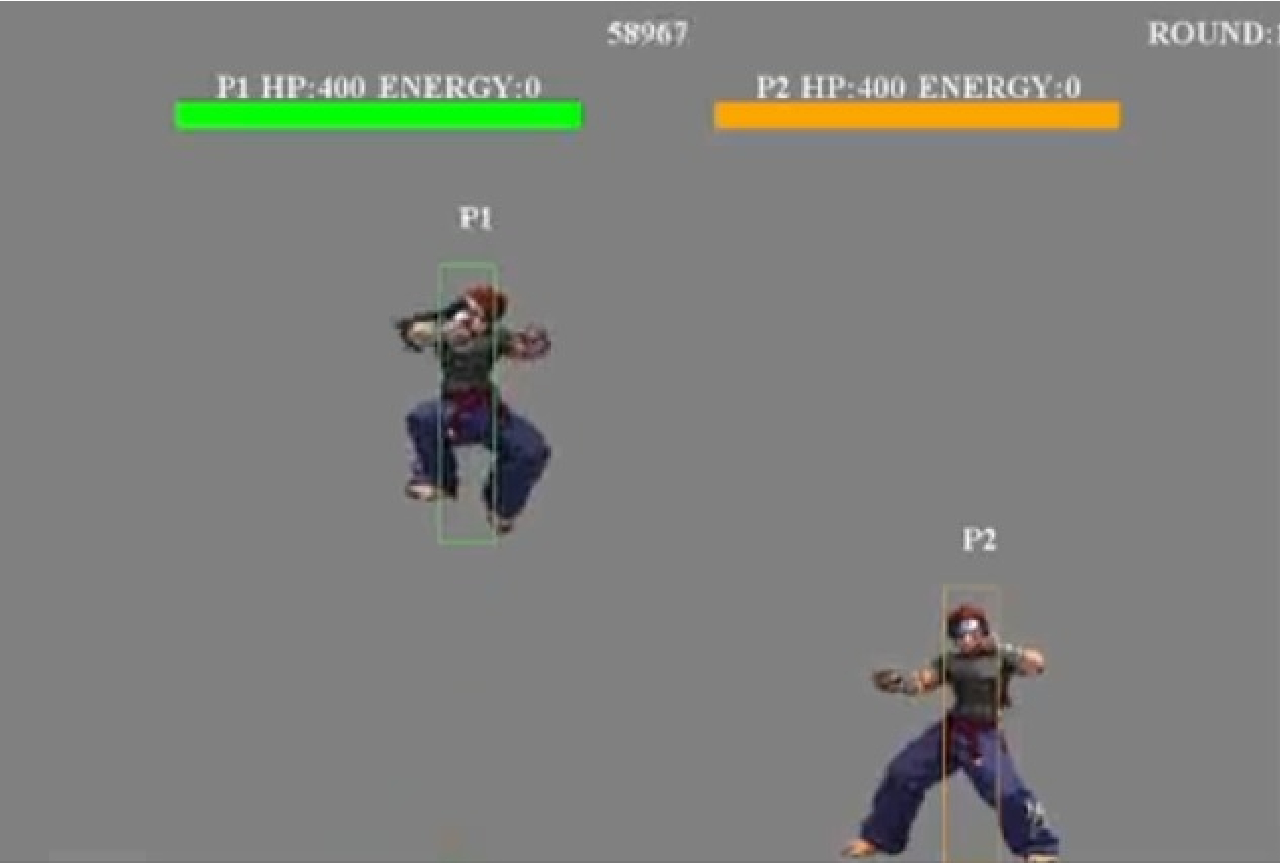}\hfill
    \includegraphics[width=.33\linewidth]{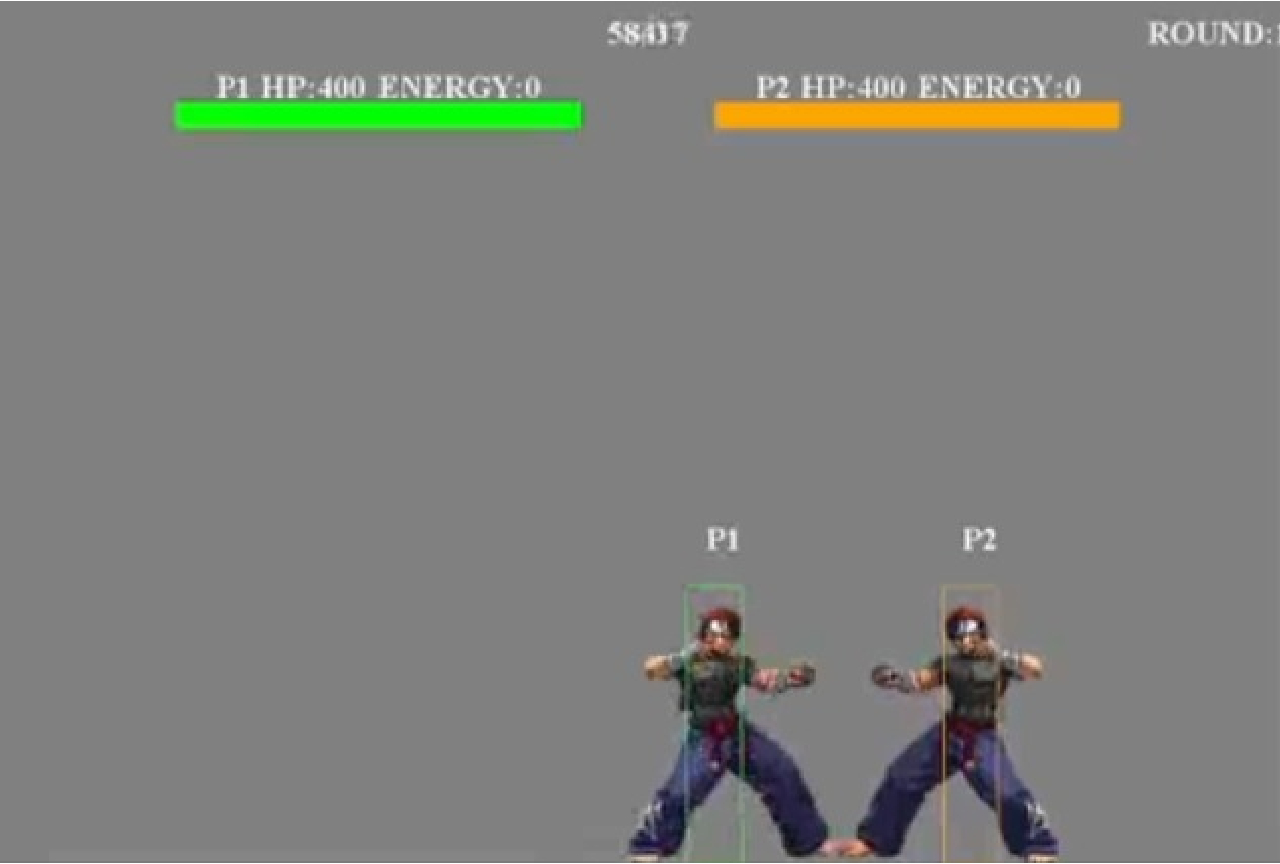}\hfill
    \includegraphics[width=.33\linewidth]{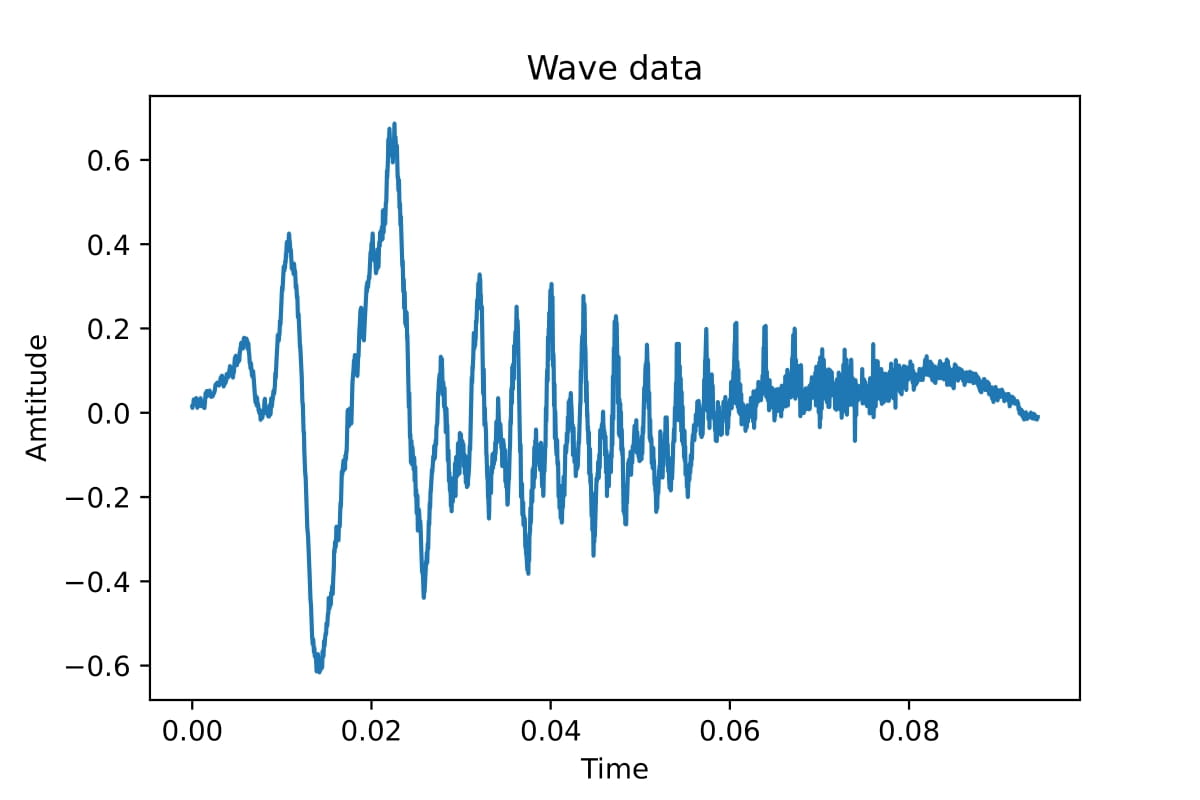}\hfill
    \includegraphics[width=.33\linewidth]{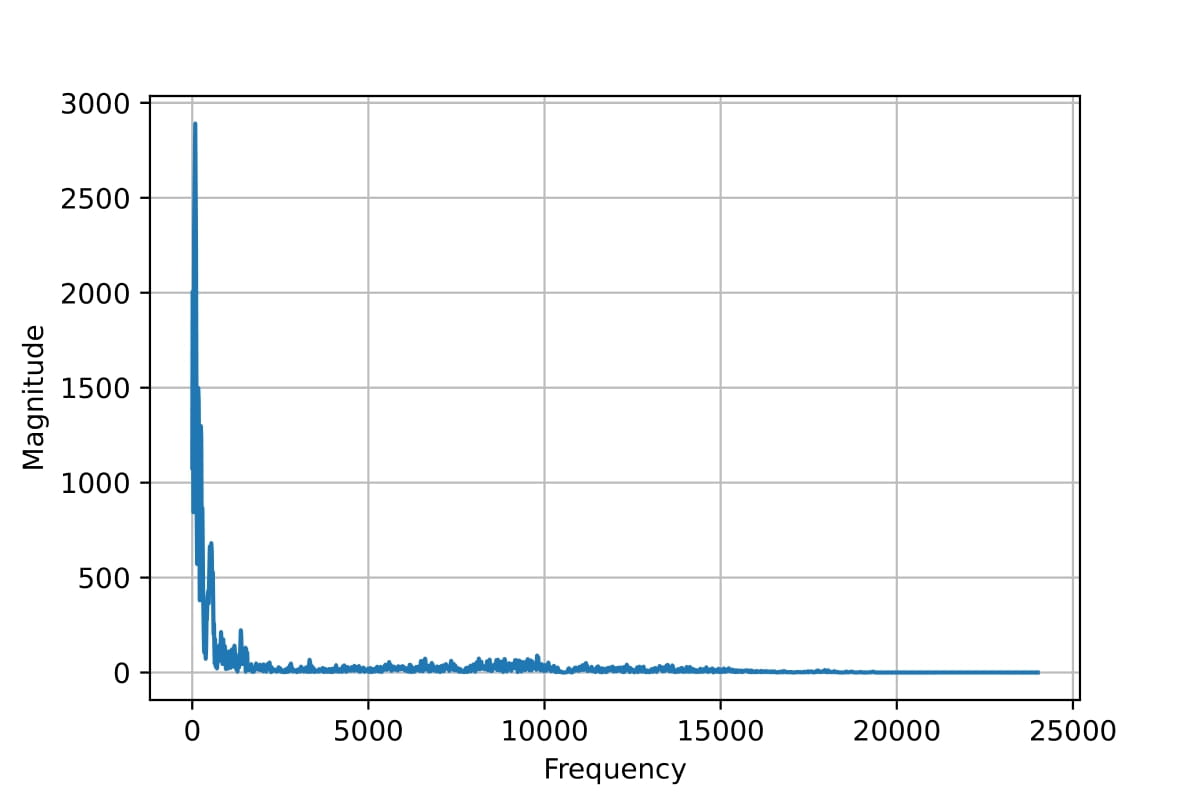}\hfill
    \includegraphics[width=.33\linewidth]{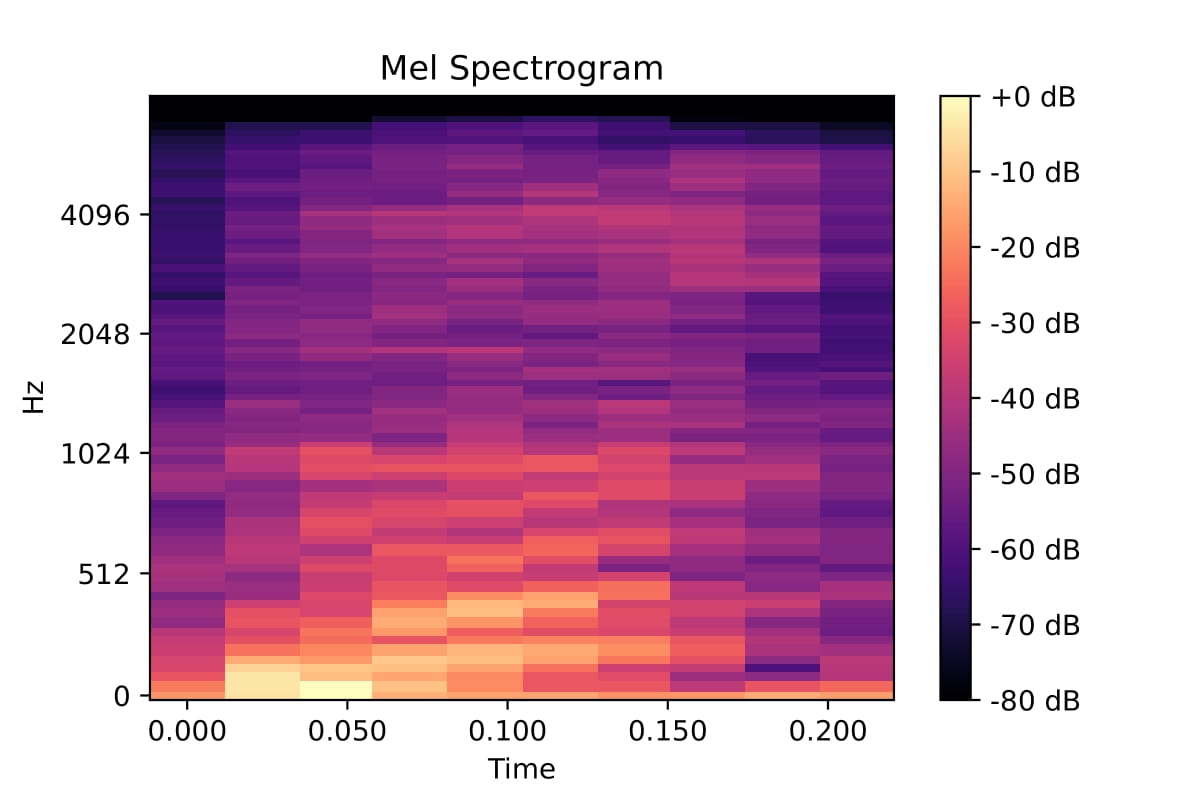}\hfill
    \caption{JUMP action in DareFightingICE. The three in the top show the action sequence of ZEN (the official character in the 2022 competition), the others below show raw sound data, Fourier Transform and Mel-Spectrogram, respectively.}
    \label{fig:figAudioProcessSample}
\end{figure*}

As shown in Fig. ~\ref{figSoundDataStructure}, not only are raw data  provided at each frame, we also provide two transformations of data (Fig.~\ref{fig:figAudioProcessSample}): Fast Fourier Transform (FFT) and Mel-Spectrogram (Mel-Spec).
\begin{itemize}
    \item Raw Data: The raw audio data is a vector $\bf{s} \in R^n, s_i \in [-1, 1]$ where $n$ is the number of normalized audio samples. A higher sampling frequency provides more details of the observations. In our work, we perform sampling at a rate of 48000Hz, which allows for high-quality sound and fast sound rendering. In addition, sound data are sampled with a length of 16.67 ms so that our raw audio data consist of 800 samples for each of the two channels: left and right. Note that in our implementation, we perform padding so that raw data of each channel are represented by a 1024-length array for computational efficiency in the following data transformations. 
    \item FFT: We perform the first transformation from raw data to frequency domain using FFT. The output of FFT data consists of two parts: real part and imaginary part. Each part is a 1024-length array.
    \item Mel-Spec: This type of transformation has gained success in speech processing\cite{b26}. Therefore we transform raw audio samples into frequency domain spectrogram with short-term Fourier transform (STFT). STFT is a sequence of Fourier transform of a windowed signal, where the window is moved with a given hop. We then apply Mel scale to the frequencies. We use all the parameters from \cite{b26}, with a hop size of 10 ms and 80 mel-frequency components, except for the window size. Because the time length of a frame is shorter than 25 ms, we choose the window size of 1024. The output of Mel-Spec is a (80, 3) shaped vector, where 80 is the number of Mel bins and 3 is the number of Mel spectra.
\end{itemize}
 Figure ~\ref{figSoundDataStructure} shows the structure of sound data and dimension of each data type provided to the AIs. The number ``2" in each dimension represents two channels. We process both left and right audio and store the transformation data into two arrays, each of which stores the output of a single transformation. Figure ~\ref{fig:figAudioProcessSample} shows an example of audio wave and transformation of an action using our parameters.\par
 
\begin{figure}[b!]
\begin{lstlisting}
package aiinterface;
import struct.*;
/**
 * The interface that defines the methods to implement in AI.
 */
public interface AIInterface {
	 /**
	 * Gets the audio information in each frame.
	 * @param ad
	 * 			the audio information.
	 *
	 */
	default void getAudioData(AudioData ad){};
	// Other methods which are from the original FightingICE platform...
	}
\end{lstlisting}
\caption{The AIInterface interface}
\label{figAIInterface}
\end{figure}
AIs are written in Java or Python and must implement the AIInterface interface shown in Figure~\ref{figAIInterface}. The new method $getAudioData$() provides sound data at each frame of the game. The AI is provided with an $AudioData$ object as the argument to this method. $AudioData$ provides access to information about the aforementioned sound data (Raw Data, FFT, and Mel-Spec).

Our sample blind AI, which uses the interface and sound data provided by DareFightingICE, learns to play the game from sound only. Many Reinforcement Learning techniques have been used to train AIs such as Deep Reinforcement Learning in the work of Mnih et al.\cite{b27} and Proximal Policy Optimization\cite{b28} used in the AI of Rongqin Liang et al.\cite{b29}. Therefore, in our work, we use Reinforcement Learning to train our sample AI. The higher the quality sound design is, the better it represents the game environment, and, therefore, the AI can be expected to better learn the observation provided by the sound data. Thereby, the efficiency of learning progress can be used to evaluate the quality of each submitted sound design. An auxiliary competition paper for CoG 2022 on this AI is under preparation.

\section{DareFightingICE Competiton}
DareFightingICE Competition is a fighting game competition for promoting the sound design of fighting games, targeting VI Players. The competition has two tracks. The first track is a sound-design competition, and the second track is an AI competition. Participants are invited to submit a sound design for VI Players and/or an AI capable of operating with only sound input information.\par

\begin{figure}[t!]
\centerline{\includegraphics[width=0.5\textwidth]{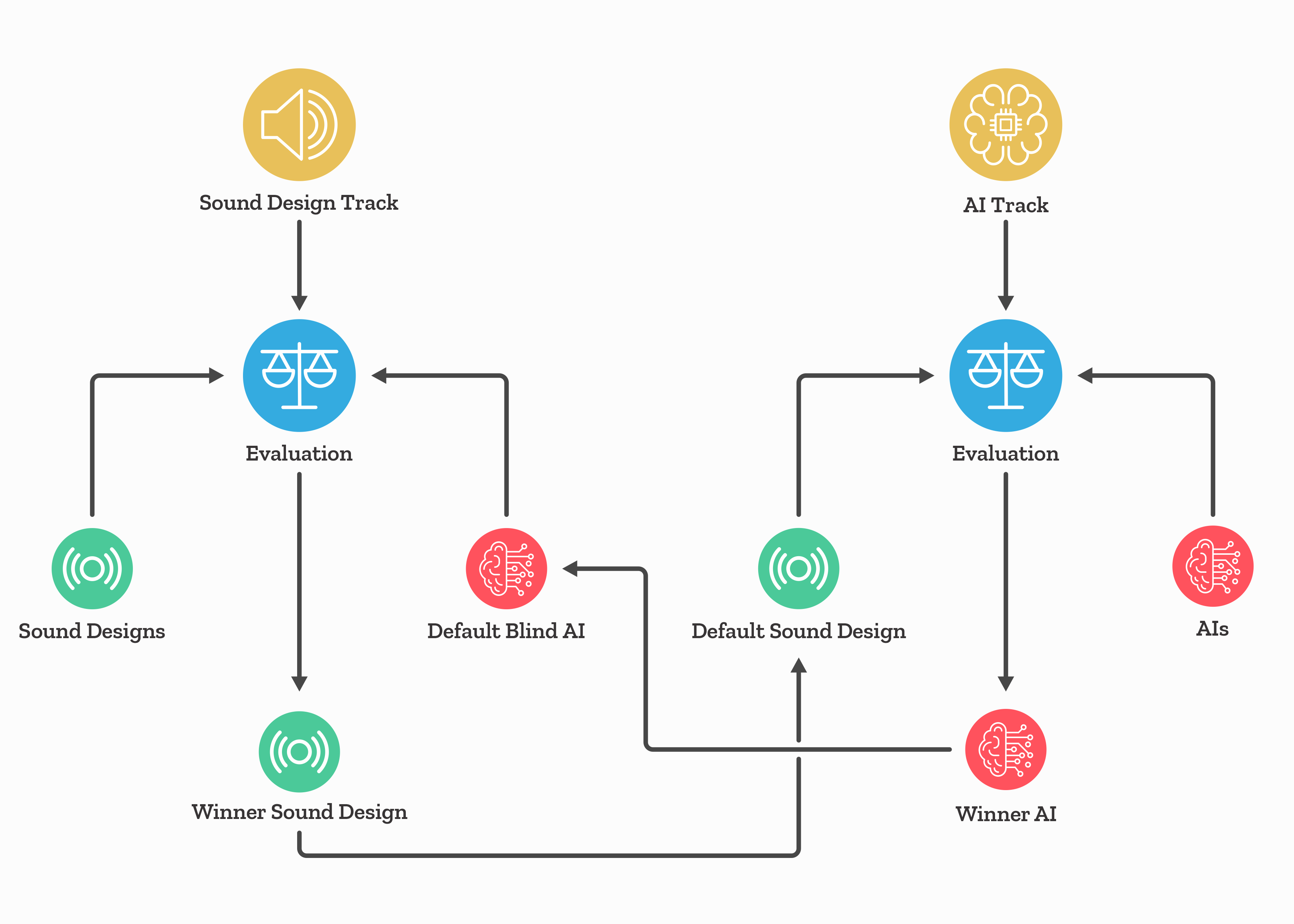}}
\caption{Competition Work Flow.}
\label{figCompWorkFlow}
\end{figure}

To achieve synergy between the two tracks, they are uniquely connected with each other. Figure~\ref{figCompWorkFlow} illustrates this point. Namely, the wining sound design of Sound Design Track will be used as the default sound design for AI Track next year. Conversely, the winner AI of AI Track will be used in Sound Design Track instead of the sample blind AI, described in IV-A and in the following. This process will continue as long as the competition is running. \par

\subsection{Sound Design Track}
In this track, participants are asked to make a sound design for DareFightingICE targeting VI Players. They are provided with a fully functional sample sound design and all the sound effects used in the sample, the default sound design of DareFightingICE. They will be allowed to edit a part of the source code in the game as well as add their own sound and any new source code files. This gives them the full ability to make their own sound design.\par

The participants are allowed to create new sound effects on their own or use already existing sound effects that have the common creative license 0\footnote{https://creativecommons.org/share-your-work/public-domain/cc0/}. Apart from creating their own sound effects, participants are allowed to use any techniques to generate sound effects or music for their sound design, including Procedural Content Generation (PCG) \cite{b30}. PCG has been used in video games to generate different content including sound effects and music \cite{b31}. Giving participants the freedom to use techniques like PCG is our way of making sure they have the resources to create a unique sound design.\par

Evaluation of sound designs will be done by players with vision while wearing a blindfold. Previous research has shown that when VI Players are not available for evaluation or testing, players with vision with blindfolds can represent them \cite{b32}\cite{b33}\cite{b34}. It was also observed that while the blindfolded players can represent VI Players, they cannot play the game as well as VI Players while being blindfolded. This is because players with vision are not used to playing with no vision \cite{b34}\cite{b35}. It is for this reason that the blindfolded players will play against a relatively weak AI in the evaluation process, but in future we plan to also add VI judges in the competition.\par

The total number of players testing a sound design is set to be a least three. If the total number of sound designs is more than five, the sound designs will go through a pre-screening process, due to the time constraint in the competition schedule, and the top five will be selected. They will then go through the screening process. If the number is five or lower, the pre-screening process will be skipped.\par

Pre-screening:

\begin{itemize}
    \item Two most capable test players (players with vision) will play against each other for one round using the default sound design. The replay of the same round will be played using all the submitted sound designs – all videos will be of the same match but different sound designs. Then a sound aesthetic survey \cite{b36} of those replay videos will be conducted targeting general respondents. The result of this survey will determine the top five sound designs.
\end{itemize}

Screening:

\begin{itemize}
    \item The blindfolded players will play against the aforementioned weak AI, a sample Monte-Carlo tree search (MCTS) AI of FightingICE \cite{b22}). Note that this weak MCTS AI will play in a non-blind fashion, having access to all the game data provided by the game.
    \item Each player will play against the AI for three games\footnote{There are three rounds per game, with an initial health point (HP) of 400 and the maximum round time of 60 seconds. Henceforth, unless stated otherwise, this setting is applied to all of the games in this work.} for each sound design, and the score will be calculated by the HP difference between the player and the AI, in relative to the HP difference when playing without being blindfolded.
    \item After the play, they (the players) will be asked to complete the same sound aesthetic survey mentioned above. The results of this survey will also be counted in the final score of a sound design of interest.
    \item In addition, our deep reinforcement learning blind AI will be newly trained with each sound design and then play the game (30 games or 90 rounds per sound design) against the same weak MCTS AI.
    \item The sample blind AI's learning curve, win-lose ratio, and HP difference will also be used in the total score. In the end, the sound design with the highest overall score will win this track. 
\end{itemize}

\subsection{AI Track}
In this track, participants will be asked to make an AI that plays DareFightingICE using only in-game sound data. Participants are allowed to submit their AIs written in Java or Python. The source code of our weak MTCS AI and sample blind DL AI will be provided. 

Following the style of the Fighting Game AI Competition run at CIG/CoG during 2014-2021, there are two leagues in this track:
\begin{itemize}
    \item The Standard League considers the winner of a round as the one with the HP above zero at the time its opponent's HP has reached zero. Both AIs will be given an initial HP of 400. The league is conducted in a round-robin fashion with two games for any pair of entry AIs switching P1 and P2, the players who start from the left side and the right side, respectively. The AI with the highest number of winning rounds becomes the league winner; if necessary, the remaining HPs are used for breaking ties. In this league, our weak sample MTCS AI, playing in the non-blind mode, and our sample blind DL AI, playing in the blind mode, will also participate as guests.
    \item In the Speedrunning League, the league winner is the AI with the shortest average time to beat both of our aforementioned sample AIs. For each entry AI, five games are conducted with the entry AI being P1 and a sample AI being P2, and another set of five games with the entry AI being P2 and the sample AI being P1. If a sample AI of interest cannot be beaten in 60 seconds, the beating time of its opponent entry AI will be penalized to 70 seconds.
\end{itemize}
\subsection{Environments}
In DareFightingICE Competition, a single environment will be used to run both tracks. More specifically, several computers will be used that have the same specification, i.e., CPU: Intel(R) Xeon(R) W-2135 CPU@ 3.70GHz 3.70 GHz, RAM: 16 GB, GPU: NVIDIA Quadro P1000 4GB VRAM, and OS: Windows 10. These computers will be deployed to evaluate sound designs submitted in Sound Design Track and to assess AIs submitted in AI Track.

\section{AI Selection For DareFightingICE}

This section describes AI vs. AI and Human vs. AI experiments. In the AI vs. AI experiment, 
three MCTS AIs with different parameter settings are used to find a suitable opponent AI for not only as an opponent AI fighting against the blindfolded players in Sound Design Track but also as one of the guest AIs in AI Track. These three MCTS AI named MctsAI165, MctsAI115, and MctsAI65 are the same as a sample MCTS AI\footnote{http://www.ice.ci.ritsumei.ac.jp/~ftgaic/Downloadfiles/MctsAi.zip} 
from FightingICE, but with the MCTS execution time \cite{b37} set to 16.5 ms (the original setting), 11.5 ms, and 6.5 ms, respectively. The execution time sets the time budget of the MCTS algorithm in use, and hence the strength of MCTS will theoretically drop when the execution time decreases.

\subsection{AI vs. AI}
This experiment evaluates performance among the three MCTS AIs. Each MCTS AI played 50 games against each of the other two AIs, switching sides between P1 and P2 every 25 games. The ratio of the number of winning rounds over 300 rounds is then used to evaluate the performance of each MCTS AI. The winning ratios of MctsAI165, MctsAI115, and MctsAI65 are 0.96, 0.49, and 0.05, respectively.

\subsection{Human vs. AI}
The experiment here examines human players' performance against the three MCTS AIs in normal play (without being blindfolded) and against the selected MCTS AI in blindfolded play. Three human players participated in this experiment and they played against the three MCTS AIs in different orders, according to the Latin square order. The human players always played as P1 three games against a given opponent AI.\par

We define $\delta$ \eqref{eqUnblindfold} to measure the normalized performance of a player of interest against an opponent AI, where $d$ is given in \eqref{eqBasic}, and due to the initial setting of HP, $min$ and $max$ are -400 and  400, respectively. A value near 0.5 of $\delta$ indicates that both player and AI have a similar fighting performance. Finally, we define ``Performance Retention Rate (PRR)" in \eqref{eqBlindfold} to show how much the players' performance is retained in blindfold play.\par

\begin{equation}
\delta=\frac{d-min}{max-min}\label{eqUnblindfold}
\end{equation}
\begin{equation}
d=\rm{HP}_{Player}-\rm{HP}_{AI}\label{eqBasic}
\end{equation}
\begin{equation}
\rm{PRR} = \frac {\rm{average}(\delta _{blind})}{\rm{average}(\delta _{normal})}\label{eqBlindfold}
\end{equation}

\begin{table}[t!]
\caption{Normalized performance of each player in normal play against different AIs, where x and y in x (y) show the average and standard deviation, respectively}
\label{tblUnblind}
\begin{center}
\begin{tabular}{|c|c|c|c|}
\hline
\textbf{\diagbox{Player}{AI}} & \textbf{MctsAI165}& \textbf{MctsAI115}& \textbf{MctsAI65} \\
\hline
Player 1& 0.22\:(0.11)&0.34\:(0.11)&0.43\:(0.15)  \\
\hline
Player 2& 0.23\:(0.08)&0.36\:(0.11)&0.59\:(0.08)  \\
\hline
Player 3& 0.44\:(0.12)&0.46\:(0.16)&0.51\:(0.21)  \\
\hline
\rowcolor{gray!20}
Average & 0.30&0.39&0.51\\
\hline
\end{tabular}
\end{center}
\end{table}

\begin{table}[t!]
\caption{Normalized performance and PRR of each player in blindfold play against the selected MctsAI65, where x and y in x (y) show the average and standard deviation, respectively}
\label{tblBlind}
\begin{center}
\begin{tabular}{|c|c|c|}
\hline
Player & $\delta$ & PRR\\
\hline
Player 1& 0.26\:(0.09) & 0.59\\
\hline
Player 2& 0.46\:(0.05) & 0.78\\
\hline
Player 3& 0.38\:(0.12) & 0.74\\
\hline
\rowcolor{gray!20}
Average & 0.36 & 0.71\\
\hline
\end{tabular}
\end{center}
\end{table}

Table \ref{tblUnblind} shows that the human players have a higher performance when the execution time of MCTS AI decreases in normal play. On average, they have close fights with MctsAI65, as shown by 0.51 on the bottom-right cell. Since fighting against a too strong opponent AI such as the other two AIs would make the resulting PRR unreliable, MctsAI65 is selected as the aforementioned weak MCTS AI.\par

Table \ref{tblBlind} shows that on average PRR against MctsAI65 is 0.71. This indicates that the players have a mild performance drop and can keep some of their playing skills with the default sound design\footnote{Video showing a game in blind play: https://tinyurl.com/cog2022blind}. However, a better sound design should lead to a higher value of PRR. As a result, PRR is used as one of the metrics to evaluate sound designs in Sound Design Track.\par

\section{Conclusions and Future Work}
In this paper, we introduced our new DareFightingICE Competition which has two tracks: Sound Design Track and AI Track. We also outlined the format and the motivation behind this competition. The competition is set to run at CoG 2022. The platform for the competition is DarefightingICE -- an enhanced version of FightingICE -- with an enhanced sound design and a new AI interface that can give sound data to AIs.

For evaluation of submissions to this competition, we conducted an experiment to find the best suited opponent AI that our blindfolded human players will play against in Sound Design Track. The experiment was divided into two parts: AI vs AI and AI vs Human. Different variations of an MCTS AI were used, and it was found that MctsAI65 was best suited for the evaluation process of this track.  Playing against MctsAI65 in blindfold play led to an average PRR of 0.71, and we perceived that a better sound design should at least have a PRR higher than 0.71.

In the future, we plan to strengthen our sample blind DL AI by taking a new or different approach, such as multimodal transfer reinforcement learning \cite{b38}. We also plan to create a sound design using PCG for DareFightingICE and to continue running DareFightingICE Competition at subsequent CoG for a number of years to ensure that our participants get enough time and resources to come up with better and better solutions. Finally, findings from this competition will be regularly summarized and reported by us at subsequent CoG or journals, such as IEEE Transactions on Games.




\vspace{12pt}


\begin{thebibliography}{00}
\bibitem{b1} J. Clement, ``Video game industry - Statistics \& Facts," 2021. [Online]. Available: https://www.statista.com/topics/868/video-games/\#dossier-chapter1. [Accessed: 28 Feb., 2022].
\bibitem{b2} G. N. Yannakakis, N. Georgios  and J. Togelius, \emph{Artificial intelligence and games}, New York: Springer, Vol. 2, 2018.
\bibitem{b3} Q. Yin, J. Yang, W. Ni, B. Liang, and K. Huang, ``AI in Games: Techniques, Challenges and Opportunities," arXiv preprint arXiv:2111.07631, 2021.
\bibitem{b4} I. Khaliq, and I. D. Torre, ``A study on accessibility in games for the visually impaired," in \emph{Proceedings of the 5th EAI International Conference on Smart Objects and Technologies for Social Good}, pp. 142-148, Sep. 2019.
\bibitem{b5} R. D. Gaina and M. Stephenson, ````Did You Hear That?” Learning to Play Video Games from Audio Cues," in \emph{Proceedings of the 2019 IEEE Conference on Games (CoG)}, pp. 1-4, 2019.
\newtildeON[C]
\bibitem{b6} Feiyu Lu, Kaito Yamamoto, Luis H. Nomura, Syunsuke Mizuno, YoungMin Lee, and Ruck Thawonmas, ``Fighting Game Artificial Intelligence Competition Platform," Proc. of the 2nd IEEE Global Conference on Consumer Electronics (GCCE 2013), Tokyo, Japan, pp. 320-323, Oct. 1-4, 2013. (https://www.ice.ci.ritsumei.ac.jp/~ftgaic/index.htm)
\newtildeOFF
\bibitem{b7} K. Tuuri, O. Koskela, J. Vahlo, and H. Tissari, ``Identifying the Impact of Game Music both Within and Beyond Gameplay," in \emph{International Conference on Entertainment Computing, Springer, Cham}, pp. 411-418, Nov. 2021.
\bibitem{b8} J. Byun, and C. S. Loh, ``Audial engagement: Effects of game sound on learner engagement in digital game-based learning environments," in \emph{Computers in Human Behavior}, vol.46, pp.129-138, 2015.
\bibitem{b9} J. Zhang, and X. Fu, ``The influence of background music of video games on immersion," in \emph{Journal of Psychology and Psychotherapy}, vol. 5, no. 4, p.1, 2015.
\bibitem{b10} F. Andersen, C. L. King, and A. A. Gunawan, ``Audio Influence on Game Atmosphere during Various Game Events," in \emph{Procedia Computer Science}, vol.179, pp.222-231, 2021.
\bibitem{b11} C. Johanson, and R. L. Mandryk, ``Scaffolding player location awareness through audio cues in first-person shooters," in \emph{Proceedings of the 2016 CHI Conference on Human Factors in Computing Systems}, pp. 3450-3461, May 2016.
\bibitem{b12}  P. Ng, and K. Nesbitt, ``Informative sound design in video games," in \emph{Proceedings of the 9th Australasian conference on interactive entertainment: matters of life and death}, pp. 1-9, Sep. 2013.
\bibitem{b13} R. Andrade, M. J. Rogerson, J. Waycott, S. Baker, and F. Vetere, ``Playing blind: Revealing the world of gamers with visual impairment," in \emph{Proceedings of the 2019 CHI Conference on Human Factors in Computing Systems}, pp. 1-14, May 2019.
\bibitem{b14} C. Chai, B. T. Lau, and Z. Pan,  ``Hungry Cat—a serious game for conveying spatial information to the visually impaired," in \emph{Multimodal Technologies and Interaction}, vol. 3, no. 1, p.12, 2019.
\bibitem{b15} J. Walton-Rivers, P. R. Williams and R. Bartle, ``The 2018 Hanabi competition," in \emph{Proceedings of the 2019 IEEE Conference on Games (CoG)}, pp. 1-8, 2019.
\bibitem{b16} N. Justesen, L. M. Uth, C. Jakobsen, P. D. Moore, J. Togelius and S. Risi, ``Blood Bowl: A New Board Game Challenge and Competition for AI," in \emph{Proceedings of the 2019 IEEE Conference on Games (CoG)}, pp. 1-8, 2019.
\bibitem{b17} S. Reis, L. P. Reis and N. Lau, ``VGC AI Competition - A New Model of Meta-Game Balance AI Competition," in \emph{Proceedings of the 2021 IEEE Conference on Games (CoG)}, pp. 01-08, 2021.
\bibitem{b18} Q. Tyrell Davis, ``Carle's Game: An Open-Ended Challenge in Exploratory Machine Creativity," in \emph{Proceedings of the 2021 IEEE Conference on Games (CoG)}, pp. 01-08, 2021.
\bibitem{b19} Game Music and Sound Design Conference, 2018. [online] Available at: https://www.gamesoundcon.com/ [Accessed: 28 Feb., 2022].
\bibitem{b20} The Berlin International Sound Design Competition, 2018. [online] Available at: https://www.bifsc.org/sound-design-competition/ [Accessed: 28 Feb., 2022].
\bibitem{b21} Film Music Contest's Music for video game, 2021. [online] Available at: https://www.fmcontest.com/music-for-video-game/ [Accessed: 28 Feb., 2022].
\bibitem{b22} R. Ishii, K. Fujimaki and R. Thawonmas, ``Fighting-Game Gameplay Generation Using Highlight Cues," in \emph{IEEE Transactions on Games}, doi: 10.1109/TG.2021.3097071, 2021.
\bibitem{b23} T. Chen, F. Richoux, J. M. Torres and K. Inoue, ``Interpretable Utility-based Models Applied to the FightingICE Platform," in \emph{Proceedings of the 2021 IEEE Conference on Games (CoG)}, pp. 1-8, 2021.
\bibitem{b24} Z. Tang, Y. Zhu, D. Zhao and S. M. Lucas, ``Enhanced Rolling Horizon Evolution Algorithm with Opponent Model Learning," in \emph{IEEE Transactions on Games}, doi: 10.1109/TG.2020.3022698, 2020.
\bibitem{b25} S. Hegde, A. Kanervisto and A. Petrenko, ``Agents that Listen: High-Throughput Reinforcement Learning with Multiple Sensory Systems," in \emph{Proceedings of the 2021 IEEE Conference on Games (CoG)}, pp. 1-5, 2021.
\bibitem{b26} D. Garcia-Romero, G. Sell, and A. Mccree, ``Magneto: X-vector magnitude estimation network plus offset for improved speaker recognition," in \emph{Proceedings of the Odyssey 2020 the speaker and language recognition workshop}, pp. 1-8, May 2020.
\bibitem{b27} V. Mnih, K. Kavukcuoglu, D. Silver, A. Graves, I. Antonoglou, D. Wierstra, and M. Riedmiller, ``Playing atari with deep reinforcement learning," in \emph{NIPS Deep Learning Workshop}, 9 pages, 2013.
\bibitem{b28} J. Schulman, F. Wolski, P. Dhariwal, A. Radford, and O. Klimov, ``Proximal policy optimization algorithms," arXiv preprint arXiv:1707.06347, 2017. 
\bibitem{b29} R. Liang, Y. Zhu, Z. Tang, M. Yang and X. Zhu, ``Proximal Policy Optimization with Elo-based Opponent Selection and Combination with Enhanced Rolling Horizon Evolution Algorithm," in \emph{Proceedings of the 2021 IEEE Conference on Games (CoG)}, pp. 1-4, 2021.
\bibitem{b30}  N. Shaker, J. Togelius, and  M. J. Nelson, \emph{Procedural content generation in games}, Switzerland: Springer International Publishing, 2016.
\bibitem{b31} J. Liu, S. Snodgrass, A. Khalifa, S. Risi, G. N. Yannakakis, and J. Togelius, ``Deep learning for procedural content generation. Neural Computing and Applications," in \emph{Proceedings of the Neural Comput and Applications}, vol. 33, no, 1, pp. 19–37, 2021.
\bibitem{b32} J. F. P. Cheiran, L. Nedel and M. S. Pimenta, ``Inclusive Games: A Multimodal Experience for Blind Players," in \emph{Proceedings of the 2011 Brazilian Symposium on Games and Digital Entertainment}, pp. 164-172, 2011.
\bibitem{b33} A. Adkins, K. Kohm, R. Zhang, and N. Gustafson, ``Lost in Spaze: An Audio Maze Game for the Visually Impaired," in \emph{Extended Abstracts of the 2020 CHI Conference on Human Factors in Computing Systems}, pp. 1-6, April 2020.
\bibitem{b34} N. Mohd Norowi, H. Azman,  and  N. Wahiza Abdul Wahat, ``Apple Swipe: A Mobile Game Apps for Visually Impaired Users Using Binaural Sounds," in \emph{Asian CHI Symposium}, pp. 196-201, May 2021.
\bibitem{b35} A. M. Silverman, ``The perils of playing blind: Problems with blindness simulation and a better way to teach about blindness," in \emph{Journal of Blindness Innovation and Research}, vol. 5, no. 2, 2015. 
\bibitem{b36} M. H. Phan,  J. R. Keebler,  and B. S. Chaparro, ``The development and validation of the game user experience satisfaction scale (GUESS)," Human factors, vol. 58, no. 8, pp. 1217-1247, 2016.
\bibitem{b37} M. Ishihara, T. Miyazaki, C. Y. Chu, T. Harada, and R. Thawonmas, ``Applying and improving Monte-Carlo Tree Search in a fighting game AI," in \emph{Proceedings of the 13th international conference on advances in computer entertainment technology}, pp. 1-6, Nov. 2016.
\bibitem{b38}Silva, Rui, Miguel Vasco, Francisco S. Melo, Ana Paiva, and Manuela Veloso.``Playing Games in the Dark: An Approach for Cross-Modality Transfer in Reinforcement Learning."
\emph{Proceedings of the 19th International Conference on Autonomous Agents and MultiAgent Systems}, pp. 1260-1268. 2020.

\end{thebibliography}
\end{document}